\documentstyle[preprint,aps,epsf,psfrag]{revtex}
\renewcommand{\baselinestretch}{1.3}

\input epsf

\def\be{\begin{equation}}
\def\ee{\end{equation}}
\def\ba{\begin{eqnarray}}
\def\ea{\end{eqnarray}}

\begin{document}
\begin{titlepage}
\renewcommand{\thefootnote}{\fnsymbol{footnote}}
\renewcommand{\baselinestretch}{1.3}
\hfill UWThPh - 1996 - 41
\medskip
\vfill
\begin{center}
{\LARGE {Traversable Wormholes \\
in Geometries of Charged Shells\\}}
\medskip
\vfill
\renewcommand{\baselinestretch}{1}
{\large F. SCHEIN 
\footnote{e-mail: schein@pauli.thp.univie.ac.at\newline
\hspace*{8 pt}
$^{**}$e-mail: pcaich@pap.univie.ac.at}
and  P.C. AICHELBURG$^{**}$\\ 
\medskip
Institut f\"ur Theoretische Physik \\
Universit\"at Wien \\
Boltzmanngasse 5, A--1090 Wien, \\
Austria \\}
\end{center}
\vfill

\begin{abstract}
We construct a static axisymmetric wormhole from 
the gravitational field of two charged shells which are kept 
in equilibrium by their electromagnetic repulsion. 
For large separations the exterior 
tends to the Majumdar-Papapetrou
spacetime of two charged particles. The interior of the wormhole is a 
Reissner-Nordstr\"om black hole matching to the two shells.
The wormhole is traversable and connects to the same
asymptotics without violation of energy conditions.
However, every point in the Majumdar-Papapetrou region
lies on a closed timelike curve.
\end{abstract}
\bigskip
\hfill PACS numbers: 04.20.Gz, 04.20.Jb
\end{titlepage}

\section{Introduction}

When discussing the Reissner-Nordstr\"om - solution,
Hawking and Ellis in their book \cite{HawkingEllis}
mention "the intriguing possibility to travel to other universes passing
through the wormholes made by charges. Unfortunately it seems that one would not
be able to get back again to our universe..."

Carter, in 1966 \cite{Carter}, addressed this question by saying "Since
the complete manifold consists of an infinite chain of universes connected
successively in time by wormholes, one has apparently two possibilities:
either to regard each particle as being connected to an infinite set of
distinct universes, or else to devise some scheme whereby some of these
universes are identified with other."

The latter possibility was also considered by Morris and Thorne
\cite{MorrisThorne}, Visser \cite{Visser} and Hawking \cite{Hawking}.
However, no explicit identification was given. The problem is not as trivial
as one may think at first sight, since a solution where the wormhole
connects to the same asymptotic region can at best be axially 
(rather than spherically) symmetric.

In this note we give an explicit construction of a charged wormhole which does
connect to the same asymptotic region. The exterior spacetime is that of two
shells held in static equilibrium by their electric repulsion while the interior
is a Reissner-Nordstr\"om black hole. The transition between the exterior and
interior spacetime is achieved by introducing two shells of charged matter.
The matching can be made exact by making use of the image method
in analogy to the construction given by R.W. Lindquist \cite{lind} who
considered the time-symmetric initial value problem for Einstein-Rosen
manifolds.
 
Not only is the wormhole traversable and allows one
to travel back to the same universe, 
but closed timelike curves exist. Moreover,
no violation of energy conditions is necessary.  
However, it faces the problem of the instability of the inner (Cauchy)
horizon.

A similar construction for uncharged shells held in static equilibrium
by strings has recently be given by W. Israel and the authors \cite{schein}, 
but at the price of introducing exotic matter.

\section{Majumdar-Papapetrou solution}

The exterior field of a system of charged bodies which are held in
equilibrium by a balance between electrostatic repulsion 
and gravitational attraction is given 
by the Majumdar-Papapetrou solution
of Einstein-Maxwell equations. 
In cartesian coordinates this solution has the form
\begin{equation} \label{metric}
ds^{2}=-V^{-2}dT_+^{2}+V^2(dx^{2}+dy^2+dz^{2})
\end{equation}
where the function V(x,y,z) satisfies Laplace's equation,
\be \label{laplace}
\triangle V(x,y,z)= \left( \frac{\partial^2}{\partial x^2} +
\frac{\partial^2}{\partial y^2}+\frac{\partial^2}{\partial z^2} \right)
V(x,y,z)=0.
\ee
The fact that for the Majumdar-Papapetrou metric 
(\ref{metric}) Einstein's equations reduce
to the Laplace equation (\ref{laplace}) offers the possibility of
constructing an exact wormhole solution.
We cut out from the Majumdar-Papapetrou
spacetime the interior of
two (non-intersecting) spheres $S_i^+$ (i=1,2) and 
require the potential function
$V$ to be constant on the surfaces. 
The problem is analogous to that of finding the electric potential outside two
charged metal spheres.
Such solutions can be found for any location and radii and arbitrary values of the
potential on the spheres \cite{durand}.

In what follows we give an explicit construction for the symmetric two-body
problem. We choose the z-axis to point along the line of symmetry joining the two
spheres $S_i^+$ with radii $R$ and center them at $z= \pm d_1$. Moreover, we fix the
value of the potential function on the spheres to 
\be
V \vert_{S_i^+} = V_0 = 1+\frac{m_1}{R}.  
\ee
This choice ensures that for large distances of the two spheres
the field is that of two particles with $mass=charge=m_1$.
The image masses $m_n$ to make $V$ constant on $S_i^+$ 
have to be located on the z-axis at $z=\pm d_n$, where  
\be \label{recurs1}
d_n=d_1-\frac{R^2}{d_1+d_{n-1}} \quad (n>1) \\
\ee
\be \label{recurs2}
m_n=-\frac{m_{n-1}R}{d_1+d_{n-1}} \quad (n>1)
\ee
The resulting expression for the metric potential $V(\vec x)$, 
$\vec x=(x,y,z)$, is 
\be
V(\vec x)=1+ \sum_{n=1}^{\infty} \frac{m_n}{\vert \vec x  \pm \vec d_n \vert}.
\ee
Following the work of R. Lindquist \cite{lind}
we define a new pair of parameters $c, \mu_0$ by 
\ba
R=\frac{c}{\sinh{\mu_0}} \nonumber \\
d_1=c \coth{\mu_0} 
\ea
which allow us to solve the recursion formulas 
(\ref{recurs1}) and  (\ref{recurs2}):
\ba
d_n=c \coth{n \mu_0} \quad (n\ge 1) \\
m_n= (-1)^{n+1}\frac{\sinh{\mu_0}}{\sinh{n \mu_0}}m_{1} \quad (n \ge 1)
\end{eqnarray}   	
We also introduce bispherical coordinates
\ba
\coth{\mu}=(x^2+y^2+z^2+c^2)/(2cz) \nonumber \\
\cot{\eta}=(x^2+y^2+z^2-c^2)/(2c \sqrt{x^2+y^2}) \nonumber \\
\cot \varphi = \frac{x}{y}.
\ea
In this adapted coordinate system the equation for the two 
throats $S_i^+$ simply become $\mu= \pm \mu_0$
and one computes
\be
\frac{m_n}{\vert \vec x \pm \vec d_n \vert}= (-1)^{n+1}
\frac{\sqrt{\cosh{\mu}-\cos{\eta}}}{\sqrt{\cosh{(\mu \pm 2n \mu_0)}-\cos{\eta}}}
\frac{m_1}{R}
\ee
The metric (\ref{metric}) takes the form
\begin{equation} \label{bisph}
ds^{2}_+=-V^{-2}dT_+^{2}+V^2 \frac{c^2}{(\cosh{\mu}-\cos{\eta})^2}
(d\mu^2 + d\eta^2 + sin^2 \eta d \varphi ^2) ,
\end{equation}
and the solution of our boundary value problem
$V(\pm \mu_0, \eta,\varphi)=V_0$ becomes
\begin{equation}
V(\mu,\eta,\varphi)=1+\frac{m_1}{R} \left[ 1+\sqrt{\cosh{\mu} - \cos{\eta}}
\sum_{n=-\infty}^{+\infty}(-1)^{n+1}\frac{1}{\sqrt{\cosh{(\mu+2n\mu_0)}
- \cos{\eta}}} \right]
\end{equation}   	

\section{Construction of the wormhole geometry}
Having found the exterior solution we match a Reissner-Nordstr\"om 
black hole to the interior of the spheres.
This requires the introduction of two infinitely thin
shells of charged matter at the transition surfaces $S_i^+$. The wormhole
is obtained by gluing different asymptotic regions of
one and the same extended Reissner-Nordstr\"om spacetime to the 
surfaces $\mu=\pm \mu_0$.
Hence, the metric interior to the shells has the form
\be \label{RN}
ds^{2}_-=-f(r_-)dT_-^{2}+\frac{dr_-^{2}}{f(r_-)}+r_-^2(d \vartheta ^2+sin^2
\vartheta d \varphi ^2) ,
\ee
where
\be 
f(r_-)=1-\frac{2m}{r_-}+\frac{e^2}{r_-^2} \qquad (\vert e \vert \le m)
\ee
Let us take the timelike surface $S_1^-$ defined by $r_-=RV_0$ 
lying outside the event horizon in one asymptotically flat region, 
say region I (see Fig.1), of the given Reissner-Nordstr\"om spacetime and 
cut off the asymptotically flat part.
In order to match the surface $S_1^-$ 
to the exterior region at the surface $S_1^+$ 
we have to determine the identification of points on $S_1^+$ and $S_1^-$.
Therefore we introduce a spherical
polar coordinate system $(T_+,r_+,\vartheta,\varphi)$ 
centered at $z=-d_1$ such that 
sphere $S_1^+$ is given by $r_+=R$.
(Note that we have the choice of
taking a right or left handed coordinate system pointing to the positive or
negative direction of the z-axis.) The metric (\ref{metric})
of the exterior region takes the form
\begin{equation} \label{sphcoor}
ds^{2}=-V^{-2}dT_+^{2}+V^2(dr_+^{2}+r_+^2(d \vartheta ^2+sin^2
\vartheta d \varphi ^2))
\end{equation}
We have not distinguished angular components of the coordinate patches
(\ref{RN}) and (\ref{sphcoor}) because now we identify points with equal
values of $\vartheta,\varphi$ and equal proper time $\tau$ 
on the shells $S_1^+$ and $S_1^-$, $S_1^+ \equiv S_1^- \equiv S_1$. 
Hence, the induced metric on the shell $S_1$ is 
\be 
ds^{2} \vert_{S_1}=-d\tau^{2}+(RV_0)^2 (d \vartheta ^2+sin^2
\vartheta d \varphi ^2)).
\ee
To construct a traversable wormhole we repeat this procedure 
for a surface $S_2^-$ 
in the asymptotic region II of the given extended
Reissner-Nordstr\"om spacetime lying in the causal future 
of $S_1^-$. This leads to a second shell $S_2$. 
Although the coordinate system (\ref{RN}) does 
not cover regions I and II, 
it is not necessary to explicitly write down different
coordinates which cover the whole
spacetime. By symmetry  
all results such as energy density and pressures of the
shells are valid for both.

\section{Energy density and pressures of the shells}
Consider shell $S_1$.
Denoting by $n$ the unit normal to $S_1$ (directed towards the 
Majumdar-Papapetrou region), and by $u=d/d\tau$ the shell's velocity,
the components of these vectors with respect to the different 
coordinate systems (\ref{RN}) and (\ref{sphcoor}) are given by
\ba
u_+^{\alpha} &=& V_0 \quad (1,0,0,0) 
\qquad n_+^{\alpha}=  
\frac{1}{V_0} \quad (0,1,0,0) \\
u_-^{\alpha} &=& \frac{1}{\sqrt{f(RV_0)}} \quad (1,0,0,0) 
\qquad  \qquad n_-^{\alpha}=\sqrt{f(RV_0)} \quad (0,1,0,0)
\ea
Applying the usual formalism of thin shells the stress $p$ and surface energy 
density $\sigma$ of the shell can be expressed by the
jump in the extrinsic curvature $[K_{ab}]$ on this surface \cite{BarrabesIs}:
\ba
\sigma &=& - \frac{1}{4\pi}[K^{\vartheta}_{\vartheta}]_{S_1} \\
       &=& - \frac{1}{4\pi V_0^2}
           \frac{\partial V}{\partial r_+} \vert_{S_1}
          - \frac{1}{4\pi RV_0} \left(
            1- \sqrt{1-\frac{2m}{RV_0}+\frac{e^2}{(RV_0)^2}} \right) \\
 p &=& \frac{1}{8 \pi} \left( [K^\tau _\tau] +
                            [K^\vartheta _\vartheta] \right)_{S_1}\\
  &=&\frac{1}{8 \pi RV_0} \left( \frac{1-\frac{m}{RV_0}}
                    {\sqrt{1-\frac{2m}{RV_0}+\frac{e^2}{(RV_0)^2}}} -1 \right)
\ea
The properties of the energy density and pressure can be inferred by decreasing
the mass and charge parameters $m$ and $\vert e \vert $ 
(note that $\vert e \vert \le m$ ) 
of the inner Reissner-Nordstr\"om region,
\ba
\lim_{m,e \rightarrow 0} \sigma &=& 
         -\frac{1}{4\pi V_0^2}
          \frac{\partial V}{\partial r_+} \vert_{r_+=R} =
         -\frac{(\cosh{\mu_0} - \cos{\eta})}{4\pi c V_0^2} 
          \frac{\partial V}{\partial \mu} \vert_{\mu=-\mu_0} \\
\lim_{m,e \rightarrow 0} p &=& 0
\ea
We can see that the sign of the surface energy density crucially 
depends on the sign
of the derivative of $V(\mu,\eta,\varphi)$ with respect to $\mu $, 
\be \label{series}
\frac{\partial V}{\partial \mu}\vert_{-\mu_0}
=\frac{m_1}{R} \left[\sqrt{\cosh{\mu_0} - \cos{\eta}}
\sum_{n=0}^{+\infty} \frac{(-1)^{(n+1)}\sinh{(2n+1)\mu_0}}
{\sqrt{(\cosh{(2n+1)\mu_0}- \cos{\eta})^3}} \right] 
\ee 
For $Cosh(\mu_0) \ge 3$, i.e. $\frac{R}{d_1} \le \frac13$, 
this quantity is negative on $S_1$ for arbitrary values 
of $\eta$ and positive mass parameter $m_1$ 
and hence, the energy density $\sigma$ is positive.
This can be seen easily from the fact that 
the first term of 
the infinite series (\ref{series}) is negative and the 
absolute values of the successive terms are monotonically decreasing. 
Although we were unable to determine the sign of the series (\ref{series})
for smaller 
values of $\mu_0$ analytically, we have numerically established 
the existence of a critical value  
$(\frac{R}{d_1})_{crit} \approx 9.993858$ at which 
at the inner poles ($\eta=\pi)$ of 
the spheres $S_i$
the energy density $\sigma$ changes sign. 

This proves that for $\mu_0 > \mu_{crit}$ and sufficient 
small values of the parameter $m$ and $\vert e \vert$ not only is 
the energy density $\sigma$ positive but all energy conditions 
are satisfied.

\section{Causal structure}
From the Penrose diagram of the extended Reissner-Nordstr\"om spacetime and the
schematic drawing of the exterior Majumdar-Papapetrou region (Fig.1) 
one sees that any spacelike slice which avoids the singularities 
(e.g. hypersurface $\Sigma$ in Fig.1) cuts $S_1$ and $S_2$ and connects 
two separated asymptotic regions.
Nevertheless any point in region $II$ 
can be connected by causal curves through the wormhole from any 
point in region $I$. 

An observer starting from the outside region and 
entering the wormhole through $S_1$ is able to reemerge 
at $S_2$ arbitrary far in the past. 
If the time gap resulting from the wormhole traversal is large enough 
he is able to travel back to his starting point in the 
exterior region and meet his "former self". 
In this sense the wormhole is an "eternal" time machine.

Notice that the condition of continuity of the induced 
metric on the surfaces $S_1$ and $S_2$ does not fix the identification
uniquely. There remains the possibility to introduce a constant 
but arbitrary shift in time.
Hence, one is able to arrange the wormhole construction 
in a way that for example observers freely falling through the wormhole 
along the z-axis (starting with a given initial velocity at $z=0$)
come back to their starting point in space and time.

Multiple traversable and non-traversable wormhole geometries may be obtained
by introducing additional shells in the Majumdar-Papapetrou region and
connecting them to the inner Reissner-Nordstr\"om solution.  

\section*{Acknowledgments}
We want to thank W. Israel for helpful discussions and 
S. Deser for critical reading of the manuscript and pointing out to us the
correct spelling of "Nordstr\"om". We also acknowledge support from the
FUNDACION FEDERICO.   

\begin{samepage}
\section*{Figure}
\begin{center}
\epsfxsize=4in\leavevmode\epsfbox{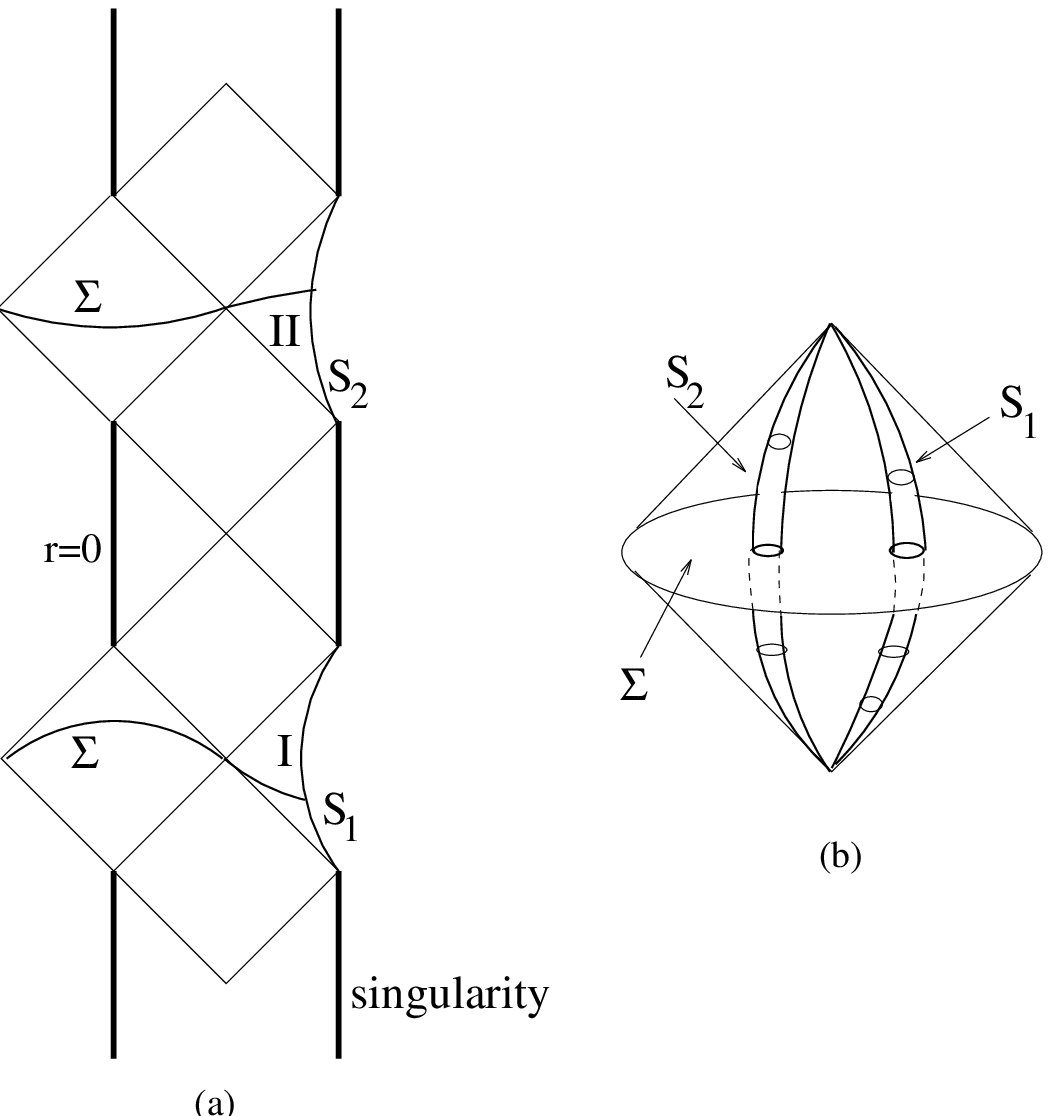}
\end{center}
\renewcommand{\baselinestretch}{1}
\small \normalsize
{\bf Figure 1:} {Wormhole geometry: (a) interior Reissner-Nordstr\"om
region, (b) schematic Penrose diagram of the exterior Majumdar-Papapetrou region} 
\end{samepage}

\end{document}